\documentstyle[12pt]{article}
\newcommand{\bea}{\begin{eqnarray}}
\newcommand{\eea}{\end{eqnarray}}
\newcommand{\nnb}{\nonumber}
\newcommand{\intD}{\mu^{4-D}\int\frac{d^Dk}{(2\pi)^D}}

\newcommand{\sla}{\!\!\!/}

\newcommand{\n}{\noindent}
\begin{document}
\thispagestyle{empty}

\vspace{0.5cm}
\begin{center}
{\large Possible large mass effects in direct determination of the
CKM elements in top decays}\\
\vspace{0.5cm}
{G. Eilam, F. Krauss and M. Lublinsky}\\
{Technion--Israel Institute of Technology}\\
{32000 Haifa, Israel}\\

%version 2.0
\vspace{0.7cm}
\parbox{10cm}{\centerline{Abstract}
\vspace{0.3cm}{\footnotesize 
We discuss the possibility that mass effects, beyond phase space,  may
have substantial influence on the direct determination of $V_{tq}$ in
decays of top quarks. In principle, our considerations are valid both for
singly produced top and for $t {\bar t}$ pair production. The mass effect
is practically irrelevant for the extraction of $V_{tb}$ from Tevatron
data, but it may have implications for higher energy multi--jet top decay. 
}}\\
\vspace{0.5cm}
\n

\end{center}

\section{Introduction}

\n Much effort is being invested, and it will intensify in the
future, in searching for New Physics beyond the Standard Model (SM).
For example, if the values of any angle of the unitarity
triangle~\cite{jarlskog} measured in two independent processes will
turn out to be different from each other, then a New Physics scenario
is unavoidable. 

Thus a thorough investigation of the elements of the
Cabibbo--Kobayashi--Maskawa (CKM)~\cite{CKM,falk} matrix is
required. Up until now the three matrix elements involving the $t$
quark, were indirectly inferred from the contributions of $t$ in
various loop processes. In addition, $V_{tb}=0.9990-0.9993$ can be
deduced~\cite{PDG} indirectly from CKM unitarity assuming the
existence of only three generations. It will be indeed exciting if
$V_{ti}$ extracted indirectly, will attain values different from those
that will be measured directly in top decays. In the present article
we discuss the possibility that mass effect, {\it i.e.} the large
disparity in mass between the $b$ and the $s$ quarks, may
significantly alter the results for $|V_{tb}|^2 / |V_{ts}|^2$ obtained
without considering the mass effect. Extra care should be therefore
practiced in carrying out the extraction of CKM elements from top
decays. 
 
Lately, the CDF collaboration has reported the measurement of the
ratio of branching ratios 

\bea
{\cal R} \equiv {\cal B}(t\to Wb)/{\cal B}(t\to Wq) 
\eea 
 
\n from $p\bar p$--collisions at the Tevatron with $\sqrt{s} = 1.8$
TeV~\cite{CDF} and deduced for the first time a direct value for  
$V_{tb}$. It was assumed that decays of the top quark to non--$W$
final states can be safely neglected. Then, taking into account that
the masses of the final state down type quarks can be ignored to high
accuracy (the relative effect of the $b$ mass on the phase space is of
the order of a few per mill), ${\cal R}$ is related to
the CKM elements of the top quark via 

\bea\label{Rdef}
{\cal R} =
\frac{\left|V_{tb}\right|^2}
{\left|V_{tb}\right|^2+\left|V_{ts}\right|^2+\left|V_{td}\right|^2}\,. 
\eea

\n Assuming three generations unitarity of the CKM matrix, the
denominator is equal to unity and thus we can identify 

\bea
{\cal R} = \left|V_{tb}\right|^2\,.
\eea

\n From the measured value ${\cal R}=0.94^{+0.31}_{-0.24}$ they
obtain~\cite{CDF} $|V_{tb}|=0.97^{+0.16}_{-0.12}$, or $|V_{tb}| >
0.75\,@\,95~\%$ CL. Now, without assuming three generations, one may
rewrite Eq.\ref{Rdef} as 

\bea\label{Vtb1}
|V_{tb}| = \sqrt{\frac{{\cal R}}{1-{\cal R}}
                \left(|V_{ts}|^2+|V_{td}|^2\right)}~.
\eea

Before discussing the main issue of the present paper, let us make a
rather trivial remark. One is tempted to use the measured ${\cal R}$
and the central values for $V_{ts}$ and $V_{td}$ as given in the
Particle Data Group tables (PDG)~\cite{PDG} to deduce
$|V_{tb}|$. However, $V_{ts}$ and $V_{td}$ were obtained using data
from the rare decays $b\to s(d)\gamma$ and from $B-{\bar B}$
mixing. But in these processes $V_{ti}~~(i=1,2)$ enter in combination
with $V_{tb}$. Therefore, in order to translate the experimental
results for the above mentioned loop processes into values of $V_{td}$
and $V_{ts}$ as in the PDG tables, three generations unitarity is
assumed. Consequently, such an approach to determine $|V_{tb}|$ from
Eq.\ref{Vtb1} is not free from the assumption of three generations
unitarity of the CKM matrix.   

In this paper we would like to dwell upon some other considerations 
related to Eq.\ref{Rdef}. The data employed to determine
$V_{tb}$~\cite{CDF} come from $t {\bar t}$ pair production at the
Tevatron. They are classified in two disjoint sets according to the
decay channels of the $W$ boson emerging in the $t\to Wq$--transition.
The final states used in the analysis are the "lepton+jets" and the
"dilepton" samples with one or both of the $W$s decaying
leptonically. In both cases the selection criteria employ cuts on the
phase space. In the following we would like to argue that such cuts on
the phase space might produce some dependence of the rates on the mass
of the down type quark in the $t\to Wq$ transition and hence spoil the
simple relation of Eq.\,\ref{Rdef}. In principle, this dependence
stems from gluon radiation. In totally inclusive calculations at
one loop order the mass effects can reach the few percent level
~\cite{Eilam}. Basically, the emission of additional gluons in
the decay of the top quarks exhibits soft and collinear divergences
that cancel when adding real and virtual contributions. However, at
every order of perturbation theory logarithmic terms remain
that depend on the relevant scales of the process, including
scales which stem from cuts on the phase space of real gluon
emissions. In double leading log approximation these logarithms can be
resummed  yielding an exponential. This result is of course the well
known Sudakov form factor~\cite{sudakov}, occurring for instance in jet
physics~\cite{ellis}. Fortunately, the mass effect has no practical
implications for the the CDF extraction of $V_{tb}$~\cite{CDF}. 

Experimentally the phase space cuts are usually related to the jet
measure definitions adopted in a given experiment. A simple and
popular definition would be of the Sterman-Weinberg type~\cite{SW},
imposing cuts on the maximal energy $\omega$ of the emitted gluons and
the angular opening of the jet $\Delta R$ (in the lab frame). In the
particular experiment~\cite{CDF}, $\Delta R$ is sufficiently large
($\Delta R \gg m_q/E_q$) and the leading double log effect is related
to $\log (\omega/E_q) \,\log \Delta R$. In this case the mass effects
are sub-leading nonexponentiable contributions of the type $m_q\, \log
(m_q/E_q)$. However, the mass effects can potentially become important
when the jet opening angle $\Delta R$ decreases toward the dead cone
value $m_q/E_q$ such that effectively the non-
-vanishing quark mass
shields the collinear divergence. In the latter case large mass
dependent logs arise. Experimentally such situations can be met in
experiments with multi--jets where more severe phase space cuts are required. 
Indeed, mass effects up to 20\% were observed in $e^+\,e^-$ annihilation 
to $b\bar b$ in three jets, and significant effects are predicted
in other environments~\cite{mjet}. Moreover, the effect
becomes more pronounced when the number of jets increases..

\section{Sudakov resummation of soft gluons}

A rigorous question of the mass effects in top decays should in
principle be addressed within a Monte Carlo jet generator approach
and will not be investigated here. In the present paper we only wish to
illustrate a potential danger of a very high jet resolution. To this
goal, let us consider an extreme (that is: unrealistic) case, imposing
simple cuts on the maximal energy of the emitted gluons only ($\Delta
R\,=\,0$). 

The squared matrix element for emission of a real gluon (momentum $k$,
$k^2 = 0$) in a $t\to Wq$ transition mediated by $\bar q\gamma^{\nu
L}t$ reads 

\bea
\left|{\cal M}_{\cal R}\right|^2 
= (-ig_s)^2 C_F 
        \left|\bar q\left[
              \gamma^\mu
              \frac{i(p\sla_q+k\sla+m_q)}
                   {(p_q+k)^2-m_q^2} 
               \gamma^{\nu L} +
               \gamma^{\nu L} 
               \frac{i(p\sla_t-k\sla+m_t)}
                    {(p_t-k)^2-m_t^2} 
               \gamma^\mu \right] t
         \varepsilon_\mu\right|^2\;,
\eea

\n which becomes

\bea
\left|{\cal M}_{\cal R}\right|^2 
=  -g_s^2 C_F \left|\bar q\gamma^{\nu L} t\right|^2 
\left(\frac{p_q^\mu}{p_qk} - \frac{p_t^\mu}{p_tk} \right)^2
\eea

\n in the soft limit. Integrating over the gluon momentum in
$D=4+2\epsilon$ dimensions, the real corrections in ${\cal
O}(\alpha_s)$ are given by 

\bea
{\cal F}_{\cal R} 
&=& -g_s^2 C_F
    \intD\,(2\pi)\delta(k^2)\theta(k_0)
    \left(
          \frac{p_q^\mu}{p_qk} - \frac{p_t^\mu}{p_tk} 
    \right)^2 \nnb\\
&=& -g_s^2 C_F\mu^{-2\epsilon} 
    \int\limits_0^\omega \frac{dk_0}{2\pi}
                         \frac{k_0^{2+2\epsilon}}{2k_0}\frac{1}{k_0^2}
         \int \frac{d\Omega_{2+2\epsilon}}{(2\pi)^{2+2\epsilon}} \\
& &   \;\;\;\;\; \times\
    \int\limits_0^\pi\,d\theta\sin^{1+2\epsilon}\theta 
         \left(\frac{\delta}{(1-\sqrt{1-\delta}\cdot\cos\theta)^2} -
               \frac{2}{(1-\sqrt{1-\delta}\cdot\cos\theta)} + 1
         \right) \nnb
\eea

\n with $\delta \equiv m_q^2/E_q^2$. Integrating over the full angular
region and bounding the gluon energy to be smaller than some maximal
value $\omega$ we find 

\bea
\lefteqn{ {\cal F}_{\cal R} = 
-\frac{\alpha_s C_F}{2\pi} 
       \left\{\frac{1}{\epsilon} \left[2 - 
 \frac{1}{\sqrt{1-\delta}} 
  \log\frac{1+\sqrt{1-\delta}}{1-\sqrt{1-\delta}} 
               \right]\right. }\nnb \\
      && \;\;\;\;\;\;\;\; +\left. \left[  
                           \frac{1}{2}   \log^2(\delta) +
      \log(\delta) 
      \log\left(\frac{\omega^2}{\mu^2}\right) \right]\right\}\,,
\eea

\n where we have given the exact result for the 
$1/\epsilon$ terms and retained only the double leading 
logarithmic terms for $\delta\to 0$ in the finite part. 
The corresponding virtual corrections to the squared matrix 
element are of the form $2\left|{\cal M}^{(1)}{\cal M}^{*(0)}\right|$,
where the matrix element to one loop order reads

\bea
2\lefteqn{{\cal M}^{(1)}
= 2(-ig_s)^2 C_F}\nnb\\
&& \cdot \intD \frac{-i}{k^2}\left[\bar q\gamma^\mu
\frac{i(p\sla_q-k\sla+m_q)}{(p_q-k)^2-m_q^2}\gamma^{\nu L} 
\frac{i(p\sla_t-k\sla+m_t)}{(p_t-k)^2-m_t^2}\gamma_\mu t \right]\,.
\eea

\n Taking into account appropriate counter terms to restore the Ward 
identities, one is therefore left with

\bea
2\lefteqn{\left| {\cal M}^{(1)}{\cal M}^{*(0)}\right|}\nnb\\
&&= ig_s^2 C_F \left|\bar q \gamma^{\nu L} t\right|^2  
         \intD \frac{1}{k^2} 
     \left(
           \frac{2p_q^\mu}{k^2-2p_qk} -
           \frac{2p_t^\mu}{k^2-2p_tk}\right)^2\,, 
\eea

\n in the soft limit $|k|\to 0$. Keeping again the exact results 
for the $1/\epsilon$ parts and the double leading logarithms 
for the finite part only, the virtual corrections are given by

\bea
\lefteqn{ {\cal F}_{\cal V} = \frac{\alpha_s C_F}{2\pi} 
               \left\{\frac{1}{\epsilon} 
\left[2 - \frac{1}{\sqrt{1-\delta}} 
\log\frac{1+\sqrt{1-\delta}}{1-\sqrt{1-\delta}}
               \right]\right.}\nnb\\
      && \;\;\;\;\;\;\;\; +\left. \left[ 
         \frac{1}{2} \log^2(\delta) +
  \log(\delta) \log\left(\frac{E_q^2}{\mu^2 }\right) \right]\right\}\,,
\eea

\n demonstrating the cancellation of soft and collinear divergences.

Resumming the real and virtual corrections and neglecting recoil
effects on the energy of the quark we find 

\bea
\exp\left[{\cal F}_{\cal{R}+\cal{V}}^q\right] = 
\exp\left[-\frac{\alpha_sC_F}{2\pi}
\log\left(\frac{m_q^2}{E_q^2}\right)
\log\left(\frac{\omega^2}{E_q^2}\right)\right]
\eea

\n for quark flavor $q$. This Sudakov form factor represents, to
double--log  accuracy, the probability that {\it no} gluon 
with energy larger than $\omega$ has been emitted during the $t\to Wq$
transition. Taking into account that $|V_{td}|$ is much smaller than
$|V_{ts}|$, Eq.\ref{Rdef} translates into 

\bea\label{Vtb2}
\frac{|V_{ts}|^2}{|V_{tb}|^2} \propto \frac{1-{\cal R}}{{\cal R}}
\cdot\frac{\exp\left[{\cal F}_{\cal{R}+\cal{V}}^b\right]}{\exp\left[{\cal
F}_{\cal{R}+\cal{V}}^s\right]} = \frac{1-{\cal R}}{{\cal R}}
\exp\left[-\frac{\alpha_sC_F}{2\pi}
\log\left(\frac{m_b^2}{m_s^2}\right)\log\left(\frac{\omega^2}{E_q^2}\right)\right]. 
\eea

\n Obviously, this ratio becomes greatly enhanced when going to more
and more exclusive measures, i.e. to lower and lower maximal energies
permitted for the gluons radiated off the quarks. For instance, if we
assume the following values 

\bea
\frac{m_b}{m_s}=45,~~~~\frac{\omega}{E_q}=0.1,~~~~\alpha_s=0.1,
\eea

\n we end up with a large enhancement factor of ${\cal O}(2)$ for the
ratio $|V_{ts}|^2/|V_{tb}|^2$.

The considered case of only soft gluon cut without imposing any cuts
on collinear gluons is certainly not physical. It merely serves us as
a calculable "toy case" to illustrate potential pitfalls in the direct
determination of $|V_{tq}|$ via top decays. For any real jet measure
the mass effect will be smaller and should be the subject of a
full Monte Carlo computation. 

\section{Conclusions}

Although we do not dispute the CDF result for $V_{tb}$~\cite{CDF},
for which a $b$ quark acts, practically, like a massless quark, 
we believe that our results show
that extra care should be practiced in extracting the CKM elements
involving the top quark. For very high jet resolutions we can get a
larger mass effect as the mass of the down type quark
decreases. This, by the way, may enhance the prospects for ``tagging'' jets
originating from light quarks. Mass effects, in particular those
present in multi--jets, have already been observed and discussed in
various environments, not including the top quark~\cite{mjet}.  
Thus, it is of essential importance for
the direct measurement of $|V_{tq}|$ to look as inclusively as
possible for $q$ quarks stemming from top decays. Finally, note that
the effect discussed here is independent of the production mechanism
of the $t$ quark, in particular whether the top is produced singly or
in a $t {\bar t}$ pair. Moreover, although the motivation of this
paper originated from the CDF measurement, our point can be of worth
to other CKM measurements especially at the next generation of 
hadron and lepton collider experiments.

\vspace{0.5cm}
{\bf Acknowledgments}

\vspace{0.2cm}
The research of GE was supported in part by the BSF, by the Israel
Science Foundation
and by the Fund for the Promotion of Research at the Technion.
FK would like to acknowledge financial
support from the Minerva--foundation.


\begin{thebibliography}{99}
\bibitem{jarlskog} C. Jarlskog, in ``CP Violation'', ed. C. Jarlskog
(World Scientific, 1988) p. 3 and references therein.
\bibitem{CKM} N. Cabibbo, Phys.Rev.Lett. {\bf 10}, 531 (1963); 
M. Kobayashi and T. Maskawa, Prog. Theor. Phys. {\bf 49}, 652 (1973).
\bibitem{falk} For a recent review of CKM elements see {\it e.g.}
A.F. Falk, {\texttt hep-ph/02011094}. See also: {\texttt http://ckmfitter.
in2p3.fr} and reference therein.
\bibitem{PDG} 
D.E. Groom {\it et al.}, Europ. Phys. J. {\bf C15}, 1 (2000) and {\texttt
http://pdg.lbl.gov}.
\bibitem{CDF} T. Affolder {\it et al.}, Phys. Rev. Lett. {\bf 86}, 3233 (2001).
\bibitem{Eilam} A. Denner and T. Sack, Nucl. Phys. {\bf B358}, 46 (1991);
G. Eilam, R. R. Mendel, and R. Migneron, Phys. Rev. Lett. 
{\bf 66}, 3105 (1991).
\bibitem{sudakov} V.V. Sudakov, Zh. Eksp. Theor. 
Fiz. {\bf 30}, 87 (1956) [JETP {\bf 3}, 65 (1956)].
\bibitem{ellis} See {\it e.g.}: R.K. Ellis, W.J. Stirling and B.R. Webber, 
``QCD and Collider Physics'', Cambridge University Press, 1996. 
\bibitem{SW} G. Sterman and S. Weinberg, Phys. Rev. Lett. {\bf 39}, 1436 
(1977). 
\bibitem{mjet} A. Ballestrero, E. Maiana and S. Moretti, Phys. Lett. 
{\bf B 294}, 425 (1992);
Nucl. Phys. {\bf B 415}, 265 (1994);
A. Ballestrero and  E. Maiana, Phys. Lett. {\bf B 323}, 53 (1994);
A. Brandenburg, W. Bernreuther and P. Uwer, Phys. Proc. Suppl. {\bf 64}, 
387 (1998);
G. Rodrigo, M. Bilenky and A. Santamaria, Nucl. Phys. {\bf B 554}, 257 
(1999); 
S. Kluth, {\texttt hep-ph/0012023};
M.L. Mangano, M. Moretti and R. Pittau, {\texttt hep-ph/0108069}.

\end{thebibliography}
\end{document}